\documentclass {article}
\usepackage{amsfonts}
\DeclareSymbolFont{AMSb}{U}{msb}{m}{n}
\DeclareSymbolFontAlphabet{\mathbb}{AMSb}
\newcommand{\complex}{\kern.1em{\raise.47ex\hbox{
            $\scriptscriptstyle |$}}\kern-.40em{\rm C}}

\input epsf
\title{Butterflies and topological quantum numbers}
\begin{document}

\author{J.~E.~Avron  and D. Osadchy
\\ Department of Physics, Technion, 32000 Haifa, Israel}
\maketitle


\input epsf
\begin{abstract}
The Hofstadter model illustrates the notion of topological quantum
numbers and how they account for the quantization of the Hall
conductance. It  gives rise to colorful fractal diagrams of
butterflies where the colors represent the topological quantum
numbers.
\end{abstract}

\section{ The Hall effect in four acts}

The first act in the Hall saga begins with a mistake made by James
Clerk Maxwell, (1831-1879). In the first edition of his book
\emph{``Treatise on Electricity and Magnetism''}, which appeared
in 1873, Maxwell discussed the deflection of a current carrying
wire by a magnetic field. Maxwell then says: \emph{It must be
carefully remembered that the mechanical force which urges a
conductor \dots acts, not on the electric current, but on the
conductor which carries it}.  If the reader is puzzled that is OK,
he should be.

In  1878 Edwin H.~Hall  was a student at Johns Hopkins University
reading  Maxwell for a class by Henry A. Rowland. Hall was puzzled
by this passage and approached  Rowland. Rowland told him that
\emph{...he doubted the truth of Maxwell statement and had
sometimes before made a hasty experiment \dots though without
success.} A schematic diagram of the scheme proposed by Rowland is
shown in Fig.~\ref{hall}. Possibly, because of this failure, Hall
made a fresh start, and tried to make accurate measurements of the
changes in the resistance---a much harder experiment. This
experiment failed, in accordance with Maxwell. Hall then decided
to repeat the experiments made by Rowland, and following a
suggestion of Rowland, replaced the original thick metal bar with
a thin gold leaf and found that the magnetic field deflected the
galvanometer needle.  This work earned Hall a position at Harvard.

Maxwell died in the year that Hall's paper came out. In the second
edition of Maxwell's book, which appeared posthumously in 1881,
there is polite footnote by the editor saying: \emph{Mr. Hall has
discovered that a steady magnetic field does slightly alter the
distribution of currents in most conductors so that the statement
in brackets must be regarded as only approximately true.} It
turned out that the magnitude, and even the sign of the Hall
voltage depends on the conductor. This made the Hall effect an
important diagnostic tool. Maxwell, even in error, inspired a
remarkable research direction.

The second act begins in 1929, with Werner Heisenberg and Rudolf
Peierls \cite{p}. As we have mentioned above, the Hall voltage was
found to be positive for some conductors and negative for others.
One sign is ``right''---it is what one would expect for (otherwise
free) electrons moving under the combined action of mutually
perpendicular electric and magnetic fields. The ``wrong'' sign was
an embarrassment. It was as if the electrons had the wrong sign
for their electric charge. The embarrassment was called the
anomalous Hall effect. Heisenberg, who pioneered the applications
of quantum mechanics to condensed matter physics, suggested to
Rudolf Peierls, who was his student at the time, to look into the
problem of the anomalous Hall effect.

Peierls, building on the results of another student of Heisenberg,
Felix Bloch, realized that the anomalous Hall effect could,
indeed, be accounted for by quantum mechanics provided  one takes
into account the periodic crystaline field.  Peierls showed that
when the conduction band is only slightly full, the electrons
behaved as if free, and the Hall response is consequently normal.
However, when the conduction band is almost full the electron go
the wrong way because of diffraction from the lattice. The
conductance turns out to be determined by the missing electrons,
i.e. the holes. The charge of a hole is opposite to the charge of
an electron and this is how Peierls resolved the anomaly \cite{p}.

The third act begins in 1980 with Klaus von Klitzing discovery
that the Hall conductance in certain two dimensional electronic
systems, in suitable ranges of experimentally controllable
parameters, is, to great accuracy, an integer multiple of $
e^2/h=(25812.807272\,\Omega)^{-1}$. This discovery led to superior
standards of resistance and to improvements in the determination
of fundamental constants. von-Klitzing was awarded the Nobel prize
in 1985 for this discovery. The theoretical developments it
spawned are wide  and deep  \cite{thouless}. In particular, it led
to the identification of the Hall conductance with a topological
invariant known as a Chern number. This scene from the third act
is going to be our main theme.

The Hall effect has a fourth act, that of the fractional quantum
Hall effect. It is no less dramatic, but, since it is not a story
of topological quantum numbers, we do not tell it.

\section{Surprising precision}\label{precision}

The precise quantization of the Hall conductance raises a puzzle.
The conductance of quantum dots is a sensitive fingerprint of the
dot. Why rearrangements of even a few atoms in a dot have
measurable consequences on the electric conductance, while in the
quantum Hall effect,  wildly different samples, manufactured in
different labs, have precisely the same quantized values for their
Hall conductance. The glib answer that the conductance of a
quantum dot and Hall conductance of two dimensional electron gas
are different, is correct, but not illuminating.

The integer Hall conductance is not only highly reproducible, it
is also a precise determination of a \emph{fundamental constant},
$e^2/h$, ($e$ is the electron charge and $h$ is Planck constant).
Why can a precision measurement of a fundamental constant be made
on macroscopic systems that are incompletely characterized? Even
more remarkable, why is the relation between the conductance and
the fundamental constant, $e^2/h$, so simple?

It is instructive to contrast this state of affairs to the
precision measurement of the (inverse) fine structure constant
$\hbar c/e^2=137.03599976$. The latter is determined from
measurements on the simplest system imaginable: a single isolated
electron. Nevertheless, the relation between what is being
measured and the fundamental constant $e^2/\hbar c$ is
complicated. One measures the anomalous magnetic moment of the
electron, $g_e+2=-0.0023193043737$, which, by quantum
electrodynamics, can be expressed as a polynomial in $e^2/\hbar
c$. The coefficients of this polynomial can be calculated in the
theory of quantum electrodynamics. This is a difficult enterprize.
The leading order of the polynomials had been calculated by
Richard Feynman and Julian Schwinger in works that vindicated
quantum electrodynamics and won them the Nobel prize in 1965. The
higher orders necessary for precision determination require
evaluating many integrals and relying on the help of computers.

In 1981 Robert Laughlin proposed a resolution of the puzzle. Here
is a variant of his  argument: Suppose one defines the conductance
as the charge added to an electrometer when the time integral of
the voltage equals the unit of quantum flux. By the principles of
quantum mechanics each measurement of the charge on the
electrometer yields an integer multiple of the electron charge.
This  quantizes the Hall conductance, simple because the number of
the electrons that one can add to an electrometer is quantized.

There is a subtle gap in this argument:  The charge on the
electrometer is only a probabilistic quantity. Which integer will
be found in any individual measurement depends probabilistically
on the wave function of the system. The measured conductance is an
average and hence need not be quantized.

To close the gap in  the Laughlin argument one needs an additional
argument why averages are also quantized. This is what topological
quantum numbers do. In the context of the Hall effect the
topological quantum numbers turn out to be the Chern numbers that
arise in the theory of fiber bundles \cite{dfn}.

\section{Topological quantum numbers} There are two
distinct ways in which physical quantities get quantized. The
familiar way one finds in textbooks of quantum mechanics is what
we shall refer to as Heisenberg quantization. An example  is the
quantization of the charge, or the number of particles, that one
finds  on an electrometer. The theoretical reason for that is that
in quantum mechanics observables are represented by matrices, and
a measurement always yields an eigenvalue of the matrix. The
operator associated with the number of particles on the
electrometer has for its spectrum the set $\{0,1,2,\dots\}$.

Topological quantum numbers are a  more arcane form of
quantization. The mechanism is different from Heisenberg
quantization. Dirac was the first to explore this avenue in his
attempt to explain quantization of charge.

Dirac addressed the fact that nature seems to have a quantum of
charge, so  the ratio of the charges of two particles is always a
rational number. For example, the charge of the proton is $-1$
times the charge of the electron. Not $-1.0001$. This is
remarkable because the electron and proton are particles whose
charges are not a-priori related in any obvious way.  For example,
their mass ratio is about $1836.109$, which is not close to a
simple fraction.

Dirac proposed a theory where the existence of a quantum of charge
was an inevitable consequence of quantum mechanics. He showed that
a magnetic monopole of charge $g$ forces the electric charge $e$
of any particle, to take value such that $\frac{2ge}{\hbar c}$ is
integral. The existence of even a single monopole in the universe
would therefore force all electric charges to be a multiple of one
basic unit of charge. For various theoretical and experimental
reasons, Dirac theory is not a completely satisfactory solution of
the charge quantization problem, but it is a paradigm for an
interesting mechanism for charge quantization and topological
quantum numbers.

The Dirac argument for the quantization of the charge is not a
consequence of the fact that the observable associated with the
product $2ge/\hbar c$ is represented in quantum mechanics by a
matrix with integer spectrum. In fact, $g$ and $e$ are treated as
ordinary numerical parameters of the theory, not as matrices.

Here is one version of Dirac quantization argument. If one tries
to write the vector potential for a magnetic monopole as a single
function throughout space one finds a singularity on a string that
terminates on the monopole. The string may be thought of as a thin
solenoid that ends at the magnetic pole and feeds the magnetic
flux. Now, if the flux carried by the string is an integral
multiple of the quantum flux unit, the singularity of the vector
potential can be removed by a gauge transformation. Since only the
modulus of the wave function (rather than its phase) and the
electromagnetic fields (rather than the potentials) have direct
physical meaning, a singularity that can be gauge away is not a
real singularity. The string is invisible to a quantum particle,
and all that remains is the magnetic monopole. If, however, the
flux is not an integer, the string can not be gauged away. It is
real, and the monopole is just a pole of a semi-infinite solenoid.

The unobservable singularity of  a Dirac string is like the
coordinate singularity of the spherical coordinate at the north
and south poles. The earth is perfectly smooth at the poles, but
the coordinates fail to be smooth there with the mildly unpleasant
consequence that there is no polar time zone.

That the Hall conductance is related to topological quantum
numbers is an observation of David Thouless, Mahito Kohmoto, Peter
Nightingale and Marcel den-Nijs \cite{tknn}.  They made this
observation for two dimensional models of non-interacting
electrons in periodic potentials. Interestingly, the topological
aspects  of these models were understood before in Boris A.
Dubrovin and Serguei Novikov \cite{novikov}. Novikov relates that
he asked his colleagues at the Landau Institute what physical
interpretation these invariants might have. Nobody gave him a
useful suggestion. It was TKNN who, independently of Dubrovin and
Novikov, identified these topological quantum numbers with the
Hall conductance.

The topological interpretation of the Hall conductance  explains
why the Hall conductance is not a fingerprint of the periodic
potential. In Figs.~\ref{hb-tb},\ref{hb-ll} this robustness can be
seen from the fact that the colored regions are open sets. In
particular, the Hall conductance does not change under small
variations of magnetic field. The Hofstadter butterfly does not
explain the quantization of the Hall conductance when
electron-electron interaction is taken into account, nor does it
explain the quantization when disorder is present.  Both play a
role in the real Hall effect. Much progress has been made in
understanding these issues, \cite{thouless,b}, but we shall not
elaborate here.

\section{Quantized averages}

Are the topological quantum numbers, and Chern numbers in
particular, really different from the ordinary quantum numbers one
is used to in quantum mechanics? To appreciate the difference
between ordinary quantum numbers and topological quantum numbers,
we look at quantum expectations.

The number operator is quantized in the sense that an individual
measurement of the number of particles in a given region always
yields an integer. However, the quantum expectation of the number
of particles need not be quantized. The quantum expectation is the
value obtained by repeated measurements on identical systems. A
peculiarity of quantum theory is that measurements are not
strictly reproducible, because the theory is not deterministic but
only probabilistic. As a consequence, even if the state of the
system is precisely specified, the outcome of a measurement may
yield different integers. Since the average of integers need not
be an integer, the average value of the number of particles need
not be quantized.

In the Dirac theory the quantization of the product $ge$ is  more
strict than Heisenberg quantization. Every measurement of $ge$
yields the same value, and not different multiples of a basic
unit. In particular, both the individual measurement and the
average are quantized and take the same value. Since both the
individual measurement and the average are quantized the
measurement is noiseless.

While the conventional Heisenberg quantization guarantees the
quantization of an individual measurement, Dirac quantization, in
the context of the Hall effect, guarantees the quantization of a
quantum expectation value.

\section{Hofstadter butterflies}

Quantum mechanics seldom leads to colorful pictures. Perhaps one
should expect this of a theory where rules for computing
probabilities replace a mental image of reality. Hofstadter
butterflies are among the few phenomena where quantum mechanics
produces colorful, fractal pictures. Besides being pretty, the
pictures also illustrate the concept of topological quantum
numbers.

Hofstadter butterflies are Escher-like diagrams of infinitely many
nested butterflies, flying to infinity.  Their monochrome version,
Fig.~\ref{hof}, was first described by Douglas Hofstadter in 1976,
in his Ph.D. work under Gregory Wannier \cite{hofstadter}.
Hofstadter  was fascinated by Mark Azbel's suggestion that under
certain circumstances, the quantum mechanical energy spectrum of
such systems can be a fractal set. Indeed, the self-similar
character of the Hofstadter butterfly turned out to be closely
related to the fractal nature of the spectrum (for irrational
values of the magnetic flux). Interestingly, the history of the
model that gives rise to the Hofstadter butterfly goes back to
Peierls who proposed it as a thesis problem to P.G. Harper.

Neither Peierls not Hofstadter considered the model in its
relation to the Hall effect, but rather as a model with intriguing
quantum mechanical spectral features. We shall take here the
opposite point of view and will not consider here the spectral
aspect of the butterfly at all. Instead we focus on the relation
of the butterfly with the quantum Hall effect.

The colored butterfly diagrams, Figs.~\ref{hb-tb} and \ref{hb-ll},
describe the electronic phases of  the quantum Hall effect. The
colors represent quantized value of the Hall conductances. Warm
colors (red) correspond to positive values for the Hall
conductance, while cold colors (blue) correspond to negative
values. White denotes vanishing Hall conductance. The quantized
values of the Hall conductances were computed using the
Diophantine equation of Thouless et.\ al.\ \cite{tknn}, (see box).
Figs.~\ref{hb-tb} and \ref{hb-ll} are the graphic expressions of
this Diophantine equation.

Fig.~\ref{hb-tb} describes the situation where the magnetic field
is the subdominant interaction. In this case, an external magnetic
field will create gaps inside a crystalline energy band. When the
Fermi energy is placed in a gaps the Hall conductance is an
integer and the gap can be assigned a color. Figs.~\ref{hb-tb}
shows the result of doing this for a large number of values for
the magnetic field. The figure repeats periodically on this axis,
with a period that is one unit of quantum flux $\frac{hc} e$. This
periodicity, is a version of Aharonov-Bohm periodicity. For
natural crystals, where the unit cell has atomic dimensions and
for the magnetic fields used in experiments on the Hall effect,
the flux through a unit cell is at most of order $10^{-4}$ of the
unit of quantum flux. This means that only a tiny sliver of the
butterfly, near the bottom of the figure, is visible for natural
crystals. A deeper exploration of the butterfly can, in principle,
be achieved by growing super-lattices with large unit cells.  The
butterfly is flanked by white margins. The white margins mean that
the Hall conductance vanishes if the (crystalline) band is either
empty or completely full. This is what Peierls expected:
Insulators should have vanishing Hall conductances.

Fig.~\ref{hb-ll}  describes the situation when the magnetic field
is the dominant interaction.  In strong magnetic fields, the
spectrum of the Schr\"odinger equation is a set of equally spaced
points, known as Landau levels.  A weak periodic potential will
broaden each of the Landau levels into a set with gaps.
Fig.~\ref{hb-ll} describes the Hall conductances when the Fermi
energy is place in the gaps. Disregarding the colors, the
butterfly then repeats periodically on the vertical axis,  with a
period that corresponds to adding a lattice cell.

 Note that the color coding is not periodic
and that while Fig. \ref{hb-tb} has inversion symmetry, the
butterflies in Fig. \ref{hb-ll} do not have this symmetry. Note
also that Fig.~\ref{hb-tb} has no albino butterflies, while
Fig.~\ref{hb-ll} does. This is one way to see that the two figures
represent different systems.

The butterfly of a broadened Landau band is an experimental
challenge because of conflicting experimental requirements
\cite{a} which were only recently overcome in  Albrecht et.\ al.\
\cite{a}.

\section{Acknowledgment} We thank A. Mann, L. Sadun, L. Schulman
and D. Thouless for useful comments. This work is supported by the
ISF and the Technion fund for promotion of research.
\section{Box: Nuts and bolts}

This box is for the reader who would like to write his own
computer program to make colored butterflies. Further information
can be found in \cite{odim}.

The are two steps involved. The first step is an eigenvalue
problem that determine the boundaries of the wings of the
butterfly. The second step is to solve a number theoretic problem
that determines an integer for each wing.  This integer is used to
color the wing.  Let us briefly outline these two steps.

The first step is to solve for the eigenvalues of a quantum
Hamiltonian describing the dynamics of an electron in two
dimensions. Using the symmetry of the problem, the two dimensional
Hamiltonian can be reduced to the study of  a one dimensional
eigenvalues problem commonly known as the Harper equation
\begin{equation}\label{harper}
\psi(n+1)+\psi(n-1)+2\cos(2\pi \Phi n)\psi(n)=E\psi(n)
\end{equation}
$E$ is the requisite eigenvalue.  In the tight binding limit,
$\Phi$ is the flux through a unit cell (in natural units). In the
limit of split Landau band, $\Phi$ is the inverse of the flux
through a unit cell.

There are no effective ways to solve this eigenvalue problem when
$\Phi$ is irrational. However, for $\Phi=\frac p q$ one looks for
periodic and antiperiodic solutions: $\psi(n+q)=\pm\psi(n)$. The
corresponding eigenvalues are then determined by two, essentially
tri-diagonal, $q\times q$ matrices. The $q$ eigenvalues of the
periodic solution determine half  the edges of the wings for that
value of $\Phi$ and the other half come from the corresponding
anti-periodic solutions.

It remains to associate a color with each band gap. The algorithm
described below is built on the original algorithm of \cite{tknn}.

Suppose that the magnetic flux through a unit cell is $\frac p q$.
For $p$ and $q$ relatively prime, define the conjugate pair
$(m,n)$ as the solutions of
\begin{equation}\label{D}
pm-qn=1
\end{equation}
$m$ is determined by this equation modulo $q$ and $n$ modulo $p$.
The algorithm for solving Eq.~(\ref{D}) is one of the oldest in
mathematics: A division algorithm of Euclid. (Standard computer
packages for  finding the greatest common divisor of $p$ and $q$,
yield also $m$ and $n$ such that $pm+qn=\gcd(p,q)$.) The Hall
conductance $k_j$, associated with the j-th gap, in the tight
binding case, is given by
\begin{equation}\label{DO}
k_j= j m \ mod \ q, \quad | k_j|\le q/2
\end{equation}
In the case of  split Landau band,  Eq.~(\ref{DO}) again
determines $k_j$ provided $p$ and $q$ are interchanged.

\clearpage

\begin{figure}
\caption{A schematics of the experimental setup of the Hall
effect. A current driven through the conductor, drawn as a prism,
leads to the emergence of voltage in the perpendicular direction.
This is the Hall voltage, which Maxwell erroneously predicted to
be zero.} \label{hall}
\end{figure}

\begin{figure}
\caption{The vector potential of a monopole is singular on a
string. When the string carries an integer number of flux quanta
it can be gauged away.} \label{dirac}
\end{figure}

\begin{figure}
\caption{Colored Hofstadter butterfly for Bloch band split by
magnetic field. The horizontal axis is the energy, or the
chemical potential; the vertical axis is the magnetic flux
through the unit cell in natural units.} \label{hb-tb}
\end{figure}

\begin{figure}
\caption{The original, monochrome, Hofstadter butterfly, describes
the spectrum of a quantum particle in a magnetic field and
periodic potential. The vertical axis is related to the magnetic
field, and the horizontal axis is the energy axis.} \label{hof}
\end{figure}

\begin{figure}
\caption{Colored Hofstadter butterfly for a single Landau level
split by a periodic potential. The horizontal axis is the energy,
or the chemical potential, the vertical axis is the
\emph{inverse} of the magnetic flux through the unit cell in
natural units.} \label{hb-ll}
\end{figure}

\end{document}